 \documentclass[final,authoryear,5p,times,twocolumn,10pt]{elsarticle}

 \usepackage{graphicx}

\usepackage{amssymb}
 \usepackage{amsthm}

\usepackage{graphics}
\usepackage{amsmath}
\usepackage[T1]{fontenc}
\usepackage[english]{babel}
\usepackage[latin1]{inputenc}
\usepackage{graphicx}
\usepackage{eepic}
\usepackage{latexsym}
\usepackage{float}
\usepackage{textcomp} 

\newcommand{\un}[1]{\mbox{ \rmfamily #1}}
\newcommand{\unp}[1]{\mbox{\rmfamily #1}}

\newcommand{\unau}[0]{\mbox{ \rmfamily au}}
\newcommand{\unpau}[0]{\mbox{\rmfamily au}}

\newcommand{\url}[1]{\ttfamily #1\normalfont}
\newcommand{\fig}[1]{Fig.~\ref{#1}}

\newcommand{\eq}[1]{Eq.~\ref{#1}}
\newcommand{\sect}[1]{Sect.~\ref{#1}}

\newcommand{\undeg}{\mbox{\textdegree}}
\newcommand{\unmin}{'}

\newcommand{\unmicron}{\mbox{\rmfamily{ \textmu m}}}

\journal{Icarus}

\begin{document}

\begin{frontmatter}



\title{Evidence for 2009 WN25 being the parent body of the November i-Draconids (NID)}


\author[NEOCC,IfA,IAPS]{Marco Micheli\corref{Phone: +39 06 941 80 365}}
\ead{marco.micheli@esa.int}
\author[IfA]{David J. Tholen}
\ead{tholen@ifa.hawaii.edu}
\author[SETI]{Peter Jenniskens}
\ead{petrus.m.jenniskens@nasa.gov}

\address[NEOCC]{SSA NEO Coordination Centre, European Space Agency,
00044 Frascati (RM),
Italy}

\address[IfA]{Institute for Astronomy, University of Hawai`i,
Honolulu, HI 96822,
USA}

\address[IAPS]{Istituto di Astrofisica e Planetologia Spaziali, Istituto Nazionale di Astrofisica,
00133 Roma (RM),
Italy}

\address[SETI]{SETI Institute,
Mountain View, CA 94043,
USA}

\begin{abstract}
In this work we propose the Amor-type asteroid 2009~WN25 as the likely progenitor of the November i-Draconids (NID, IAU\#392), a recently detected weak annual meteoroid stream. We first describe our recovery and follow-up effort to obtain timely ground based astrometry with large aperture telescopes, and ensure that 2009~WN25 would not become lost. We then discuss the possible parent-stream association, using its updated orbit to model the ejection of dust particles from the surface of the parent body and match the observed properties of the stream.
\end{abstract}


\end{frontmatter}



\section{Introduction}

The high rate of near-Earth object discoveries in the past decade resulted in a large database of objects in Earth-approaching orbits, and some of them have already been proposed as possible parents of known meteor showers. The first attempt to create a systematic list of proposed parents, published by \citet[][tables 7 and 9]{2006mspc.book.....J}, included about $60$ objects, most of which were newly proposed associations with objects in Jupiter-family comet orbits, all having orbital elements similar to the meteoroid streams. Other works, such as \citet{2008MNRAS.386.1436B}, addressed the identification of parents of specific meteor showers using less strict similarity criteria, and also proposing many asteroidal candidates.\\

If the compilation of a list of stream progenitors is not simple, the definition of a trustworthy list of meteoroid streams has been even more difficult, because historically no single organization has been responsible for the maintenance of a homogeneous census of meteor showers. The first attempt to collect all published activity reports (and the associated stream orbits) was again by \cite{2006mspc.book.....J}, who made an effort to identify as many literature references as possible to known or proposed streams. 

Starting in 2007, his work has formed the basis for a list of known showers, published and now routinely updated by Tadeusz J. Jopek at the IAU Meteor Data Center (MDC), under the authority of the International Astronomical Union (IAU). 

Even more recently, the establishment of systematic surveys such as the optical CAMS \citep{2011Icar..216...40J} and the radar CMOR \citep{2008Icar..195..317B} surveys. Nowadays, new showers are routinely established by some of these surveys, and systematically evaluated and approved by the IAU before inclusion in the official list.\\

\subsection{The November i-Draconids}

\begin{figure*}[!t]
\centering
    \includegraphics[width=0.7\textwidth]{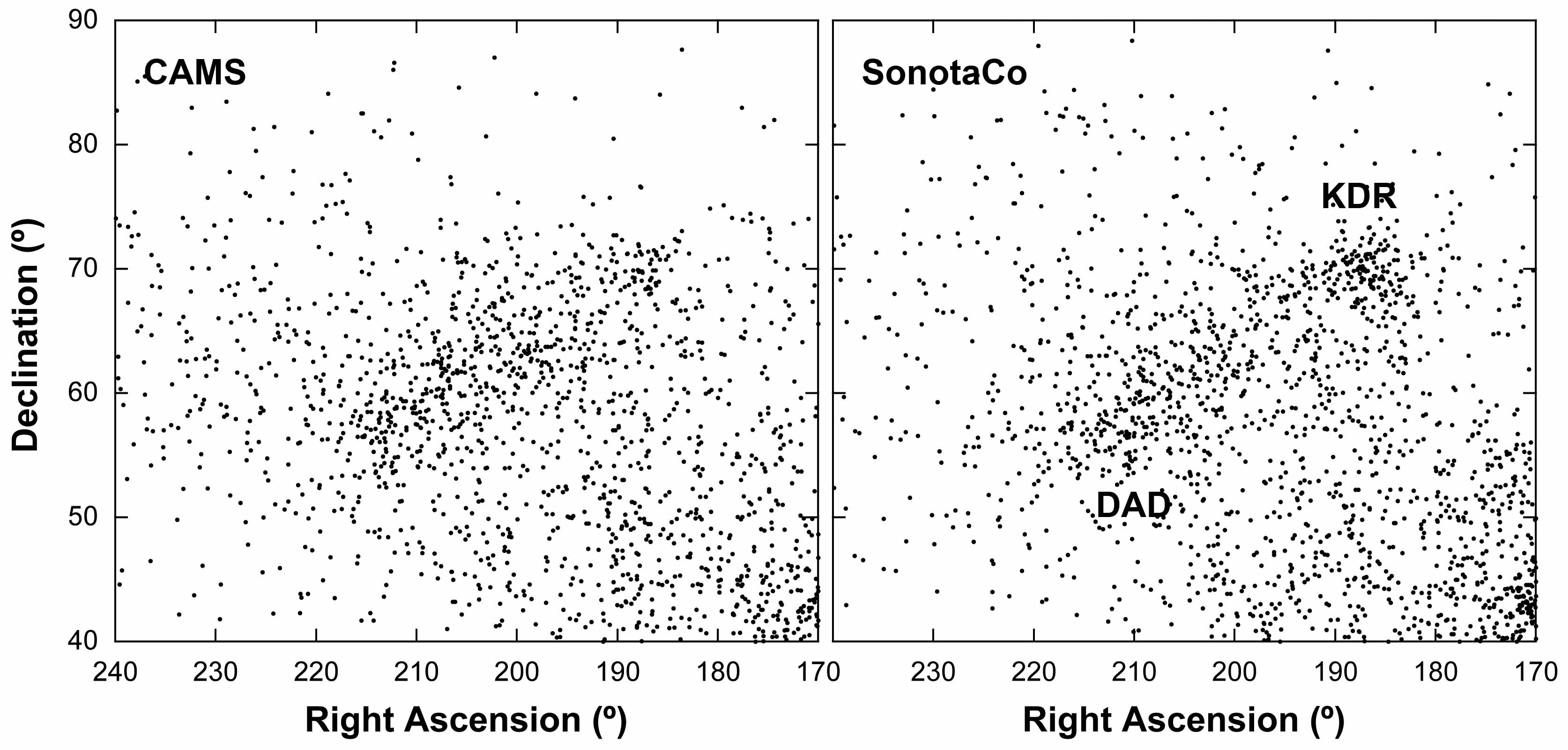}
  \caption[November i-Draconids (NID) meteors in the CAMS and SonotaCo databases]
  {Meteors in the CAMS (2010-2014) and SonotaCo (2007-2013) databases that match the required properties of the November i-Draconids (NID) (solar longitudes between $239\undeg$ and $268\undeg$), with radiant drift corrected to a solar longitude of $253\undeg$.}
  \label{NID-CAMS/SonotaCo}
\end{figure*}

The MDC list contains two established showers active in early December from the area of Draco, each describing two apparently separated clusters of activity visible in \fig{NID-CAMS/SonotaCo}; the two showers are known as the December $\alpha$-Draconids (DAD, IAU\#334), peaking at $\lambda=256.6\undeg$ from $\alpha=207.9\undeg$ and $\delta=+60.6\undeg$, and the December $\kappa$-Draconids (KDR, IAU\#336), peaking at $\lambda=250.2\undeg$ from $\alpha=186.0\undeg$ and $\delta=+70.1\undeg$.

In addition to these radiants, the same area contains an additional proposed stream with similar properties, the November i-Draconids (NID, IAU\#392)\footnote{The shower is sometimes incorrectly given as ``November $\iota$-Draconids'', with a Greek letter ``iota'' instead of an ``i''.
}, active in late November of each year from a solar longitude $\lambda \sim 241\undeg$, they originate from a radiant located at $\alpha \sim 200\undeg$, $\delta \sim +65\undeg$.

Already clearly visible in the SonotaCo Network database (\url{http://sonotaco.jp/doc/SNM/}), this complex activity was formally proposed by \cite{2010Icar..207...66B} on the basis of radar data, and it was subsequently confirmed by CAMS (see \fig{NID-CAMS/SonotaCo}). All these streams share a relatively high geocentric velocity of $v_g \sim 43 \un{km}\un{s}^{-1}$, which directly suggests a progenitor in an orbit significantly different from Earth's.

The designation November i-Draconids is now often used to identify the diffuse component that encompasses both showers, while the two established designations December $\alpha$-Draconids and December $\kappa$-Draconids identify the separate clusters visible in \fig{NID-CAMS/SonotaCo}. In the following we will use the name November i-Draconids to identify all the components together, since our results suggest that the entire complex structure of radiants can be generated by the same parent body.

\section{Establishing a possible progenitor of the stream}

A preliminary analysis of possible progenitors for the November i-Draconids, conducted according to the methods outlined in \cite{2013PhDT.......518M} and on the sole basis of the radiant information in the MDC database, shows an extremely good match of the shower's observing circumstances with those of possible stream associated with 2009~WN25, an Amor-type asteroid discovered on 2009 November 22 by the Catalina Sky Survey in Arizona, USA.
The asteroid has a very peculiar orbit, with $a \sim 3.25 \unau$, $e \sim 0.66$ and $i \sim 72\undeg$; its absolute magnitude is approximately $18.3$, suggesting a diameter of approximately $1 \un{km}$ if its albedo is at least moderately dark. 
Although inactive at the time of discovery, its Tisserand invariant with respect to Jupiter is $T_J \sim 1.96$, which immediately suggests a non-asteroidal origin and likely cometary nature.

However, at the time of this preliminary analysis 2009~WN25 only had observations spanning an arc of 25 days, from its discovery to a last detection from Siding Spring, Australia, on 2009 December 17. After that, the object became too faint, or too far south, for most follow-up stations, and would have become lost at its subsequent apparition. 
In addition, the short observational coverage, although sufficient to suggest a possible linkage with the stream, was far too short to meaningfully characterize the past dynamical evolution of the object, or to simulate possible particle ejection scenarios that may account for the generation of the observed stream. Additional follow-up was therefore necessary to confirm or discard this possible association.

\subsection{Recovery of 2009~WN25}\label{Recovery}

Around the time 2009~WN25 was first noticed as a possible stream progenitor, the object was quickly receding from Earth, with a magnitude fainter than $V=23$ and located in the Southern sky, at a declination of about $-50\undeg$. Furthermore, the skyplane uncertainty at the time was about $15\unmin$, making it a difficult target for most small-field imagers that are available on large telescopes.

\begin{figure*}[!t]
  \begin{center}
    \includegraphics[width=0.7\textwidth]{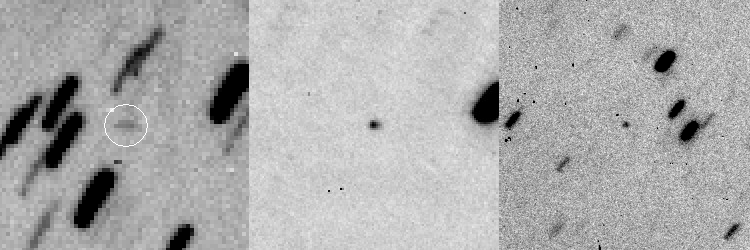}
  \end{center}
  \caption[Recovery images of 2009~WN25]
  {Left: Portion of one chip of the recovery image of 2009~WN25, obtained by Beth Biller with the $4.0\un{m}$ CTIO Blanco telescope on 2010 June 7 around 23:55 UT. The object appears slightly trailed because of the imperfect non-sidereal tracking of the telescope. The identity of the object is confirmed by other images obtained on 2010 June 6, where the object is visible but even more trailed because of an erroneous sign in the tracking rates. Center: Stack of three $100 \un{s}$ gri-filtered exposures of 2009~WN25, obtained with the $3.6\un{m}$ Canada-France-Hawaii Telescope on 2015 August 11 around 14:45 UT. At the time of these exposures, the object had magnitude $V \sim 22$, was only $22\undeg$ above the horizon, and was observed during the onset of morning twilight. The apparent extension to the left of the source does not match the expected antisolar direction of a possible tail, and it is most likely explained by chromatic aberration due to the extreme airmass and wide wavelength passband of the system. Right: Confirmation observation obtained on 2015 August 12 at 14:50 UT with the $2.2\un{m}$ University of Hawai`i reflector.}
  \label{2009WN25}
\end{figure*}

A search for possible instruments in the Southern hemisphere, with large enough aperture and field of view to detect such an object, showed that the only possible match at that time was the Mosaic II imager on the $4.0\un{m}$ CTIO Blanco telescope in Chile.
Fortunately one of the observers scheduled at the telescope around that time agreed to obtain a few exposures of the object just after twilight of her two nights at the telescope, on 2010 June 6 and 7.
These observations allowed us to recover the object about $2\unmin$ away from our own predicted ephemeris, and about $4\unmin$ from the nominal MPC solution. An image from the second recovery night is presented in the left panel of \fig{2009WN25}.

A few days later, we were able to obtain an additional single detection of the object with the $2.0\un{m}$ Faulkes Telescope South at Siding Spring, Australia. Although the detection was much fainter than the one from CTIO, it provided us with an independent confirmation that the original recovery was indeed correct.\\

These observations in 2010 already extend the observed arc on 2009~WN25 from $25$ to $203$ days, a coverage sufficient to properly model the past dynamics of the stream. Most of the results of this work can already be established on the basis of this observational arc.

However, to better understand the association it is essential that the object does not become lost even after further follow-up has been obtained. For this reason, we also attempted additional recovery observations of 2009~WN25 in May 2012, using the $8.2\un{m}$ Gemini North telescope. Observations for the program were collected under very poor seeing conditions on 2012 May 26, and the image quality was not sufficient to achieve a detection of the object, which at the time was predicted at $V=25.5$.\\

To take advantage of the next favorable observational opportunity for this object we had to wait until mid-2015, when 2009~WN25 emerged from solar conjunction at high Northern declinations, becoming observable again from the Northern hemisphere. We successfully recovered the object on 2015 August 11 with the MegaPrime one-degree imager on the $3.6\un{m}$ Canada-France-Hawaii Telescope atop Mauna Kea, Hawaii, USA (\fig{2009WN25}, central panel), and confirmed it one night later with the $2.2\un{m}$ University of Hawai`i reflector on the same mountain (\fig{2009WN25}, right panel). The object was located only $7\unmin$ away from the prediction based on our 2010 CTIO observations. Without them, the asteroid would have been more than $20\undeg$ away from the nominal position, effectively lost inside a skyplane uncertainty region more than $60\undeg$ long.

\subsection{Dust ejection models}\label{EjectionModel}

Before the association of an object with a stream can be considered significant, it is important to verify it with a dynamical simulation of the stream creation process, modeling the ejection of meteoric particles in the distant past and checking if any ejection scenario can match the current activity profile of the stream.

The first step of this modeling is to numerically integrate the orbit of the parent for a few centuries in the past. This integration is meaningful only if the orbit of the object is known with sufficient accuracy, explaining the need for the recovery observations presented in \sect{Recovery}. When a sufficient observed arc is available, ideally many months to a few years long, we can use standard integration software tools to determine the past evolution of the orbit. For this project we used the SolSyIn package\footnote{\url{http://math.ubbcluj.ro/\~{}sberinde/solsyin/index.html}} to integrate the orbit of the parent object to the desired epoch in the past, taking into account all planetary perturbers (including the Earth and the Moon as separate masses) and relativistic effects.\\

To investigate the possible creation of a meteoroid stream at a certain epoch, synthetic particles with various masses are then ejected from the object (at perihelion) with different velocities. The choice of the maximum ejection velocity is done using the empirical formula from \citet{1951ApJ...113..464W}:
\begin{equation}
v_{ej}=\sqrt{\frac{43.0~D_c }{\rho~d~r^{9/4}} - 0.559~\rho_c~D_c^2}\label{WhippleEquation}
\end{equation}
where $d$ is the size of the particle (in $\unp{cm}$) and $\rho$ is its density (in $\unp{g}\un{cm}^{-3}$), while $D_c$ and $\rho_c$ are those of the parent body (this time in $\unp{km}$ and again $\unp{g}\un{cm}^{-3}$), and $r$ is the heliocentric distance at the time of ejection (in $\unpau$, which we assume to be equal to the perihelion distance $q$). We also assume $\rho = \rho_c = 1\un{g}\un{cm}^{-3}$, meaning that the ejected particles have the same density as the parent body, assumed to be the same as water. 

In our model a few hundreds of particles of a given size are isotropically ejected from the surface of the asteroid, with velocities up to $v_{ej}$. From there on they are treated as separate bodies, and the dynamics of each one is then independently integrated forward to the present time, taking into account non-gravitational effects (such as radiation pressure and relativity) that are significant for small-size particles. 

Each particle of the cloud can now be treated as an independent meteoroid: the time of its close approach with Earth, its radiant and its entry velocity can be determined with the procedures from \citet{1998A&A...331..411N}, and compared with the observed properties of the candidate meteor shower, to confirm or reject the attribution of the stream to the specific parent that was used in the simulation. In most of these simulations there is a quite significant degeneracy between the size of the particles, the ejection velocity and sometimes the epoch of ejection. In each simulation we chose to eject particles with a range of velocities from zero to the threshold defined by \eq{WhippleEquation}, while keeping the particle size fixed. The best match between observed and simulated properties of the shower is then searched by repeating the simulation with different discrete choices of particle size and ejection time; as a result, the values presented below should be viewed only as an approximation of the actual properties of the stream.\\

Using our updated orbit of 2009~WN25 we computed a set of ejection models to attempt a match with the observed properties of the November i-Draconids stream. Different particle sizes (from $10\unmicron$ to $1000\unmicron$) and different ejection times (from $100$ to $500$ years ago) were tested, and their Earth-intersecting orbits were computed by the adjustment of perihelion method of \citet{1998A&A...331..411N} \citep[chosen because it provided the smallest orbital discrepancy as estimated by the D-criterion of][]{1963SCoA....7..261S}.

\begin{figure}[!t]
\centering
    \includegraphics[width=0.8\columnwidth]{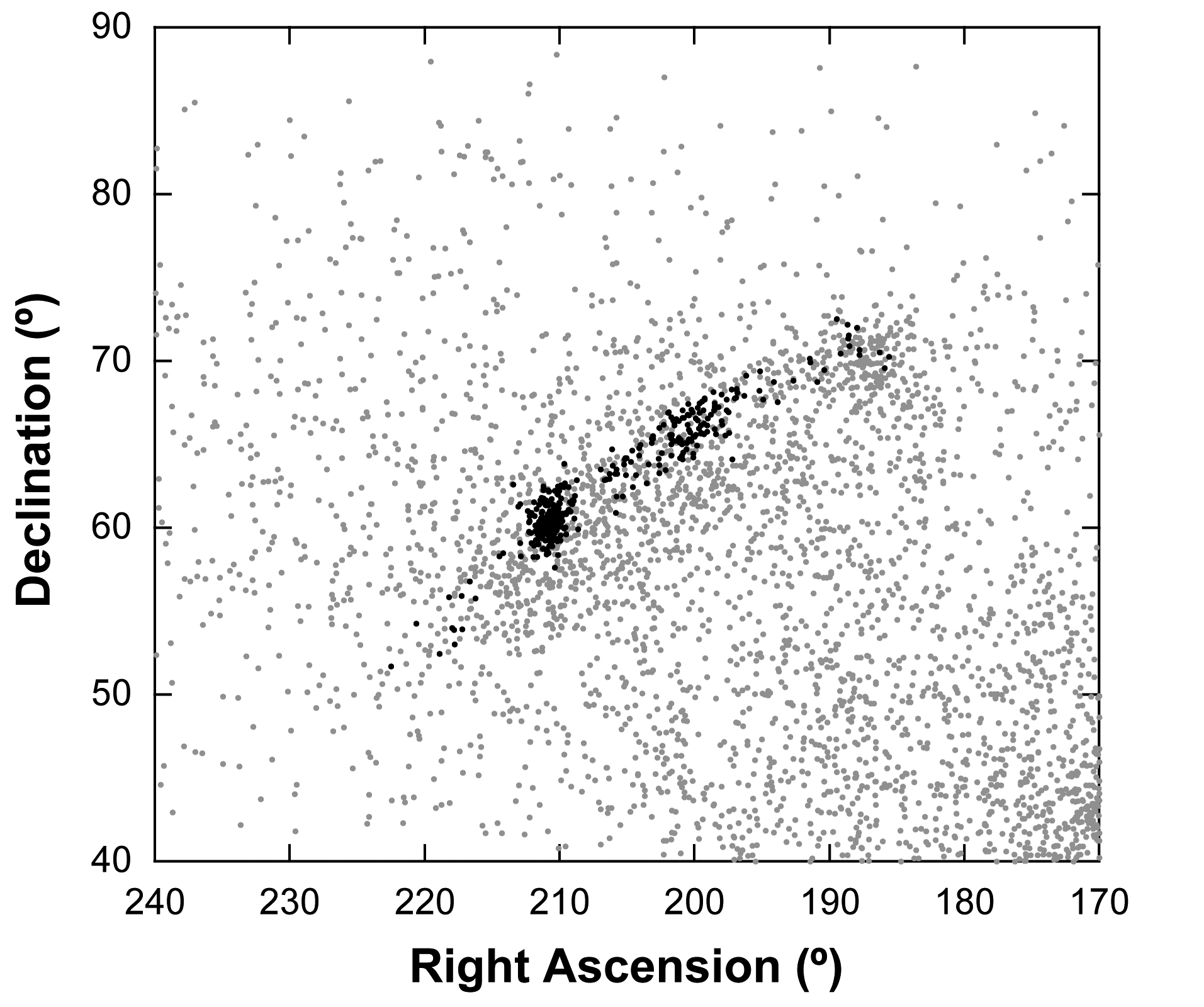}
  \caption[Ejection modeling for 2009~WN25 and the November i-Draconids (NID)]
  {Comparison of the radiant of simulated $30\unmicron$ particles ejected from 2009~WN25 $100$ years ago (black dots) with the meteors detected by the CAMS and SonotaCo surveys (gray dots, from \fig{NID-CAMS/SonotaCo}), with radiant drift corrected to a solar longitude of $253\undeg$.}
  \label{2009WN25-NID}
\end{figure}

The best possible match was obtained with particles of about $30\unmicron$ ejected from the parent body a century ago (see \fig{2009WN25-NID}), although other combinations produced similar results, the main difference being a more or less compact distribution of the radiant points around the same general structure.

The model not only replicates the appropriate radiant position, but also the complex morphology with multiple sub-radiants (DAD, NID and KDR) visible in the CAMS and SonotaCo data. However, the node of the encounter is not in agreement: the observed showers are detected during solar longitude $248\undeg$-$263\undeg$, $239\undeg$-$268\undeg$ and $250\undeg$-$255\undeg$ respectively, while the calculated clusters are encountered at solar longitude $231\undeg$-$235\undeg$, $222\undeg$-$230\undeg$ and $220\undeg$-$222\undeg$ instead.

This complex structure reproduced by our simulations is likely due to the peculiar recent dynamical evolution of 2009~WN25, which had an unusually small MOID with Jupiter for the entire last century, reaching a minimum of less than $10^{-4}\unau$ around 1971; Although the object itself did not have any exceptionally close approach around that time, coorbital particles located along the orbit could have been dramatically affected, creating the features we see in the current radiant of the November i-Draconids. Slight changes in the geometry of these close approaches may also be responsible for the mismatch of the solar longitudes noted above.

\section{Conclusions}\label{Conclusions}

In this work we presented the successful recovery of 2009~WN25, and suggested its association with the November i-Draconids (NID), a proposed complex of weak meteoroid streams. Although 2009~WN25 is currently inactive, it is shown that this NEO could have ejected meteoric particles approximately a century ago, and they would have evolved into orbits that are a good match to the observed radiant distribution of the stream. Furthermore, the peculiar structure of the November i-Draconids complex of radiants is also matched by the simulations, thus giving further credit to the association. A discrepancy in the solar longitudes of arrival may be on account of past close encounters with Jupiter.

Thanks to our observations 2009~WN25 now has a well-established orbit that will allow further monitoring in the future, instead of becoming another member of the growing collection of lost NEOs.
It is possible that 2009~WN25 may show cometary activity in one of its future apparitions, thus making evident its cometary nature and strongly supporting its association with a meteoroid stream. 
Furthermore, if taxonomic observations can be obtained in the future, they may show a primitive nature for this object, giving further support to its association to this meteor shower, even in the absence of current activity.

\section*{Acknowledgments}

The authors would like to sincerely thank Beth Biller for allowing us to use part of her telescope time at CTIO to successfully recover 2009~WN25 in 2010, and for successfully executing the observations at the telescope for us. Without her support 2009~WN25 would have certainly become lost, and this promising association would have become impossible to confirm.

Our research was funded by grants AST~0709500 and AST~1109940 from the U.S. National Science Foundation, and grant NNX12AM14G from the NASA NEOO program.

\bibliographystyle{model2-names}
\bibliography{Icar09}







\end{document}